\documentclass[12pt,preprint]{aastex}

\shorttitle{Different patterns of chromospheric evaporation}

\shortauthors{Li \& Ding}

\begin{document}

\title{Different Patterns of Chromospheric Evaporation in a Flaring Region Observed with Hinode/EIS}

\author{Y. Li and M. D. Ding}

\affil{Department of Astronomy, Nanjing University, Nanjing 210093, China;}
\affil{Key Laboratory for Modern Astronomy and Astrophysics (Nanjing University), Ministry of Education, Nanjing 210093, China}
\email{dmd@nju.edu.cn}

\begin{abstract}
We investigate the chromospheric evaporation in the flare of 2007 January 16 using line profiles observed by the EUV Imaging Spectrometer (EIS) onboard Hinode. Three points at flare ribbons of different magnetic polarities are analyzed in detail. We find that the three points show different patterns of upflows and downflows in the impulsive phase of the flare. The spectral lines at the first point are mostly blue shifted, with the hotter lines showing a dominant blue-shifted component over the stationary one. At the second point, however, only weak upflows are detected; in stead, notable downflows appear at high temperatures (up to 2.5--5.0 MK). The third point is similar to the second one only that it shows evidence of multi-component downflows. While the evaporated plasma falling back down as warm rain is a possible cause of the redshifts at points 2 and 3, the different patterns of chromospheric evaporation at the three points imply existence of different heating mechanisms in the flaring active region.
\end{abstract}


\section{Introduction}
\label{intro}

Chromospheric evaporation refers to the drastic mass motions in flaring loops caused by rapid energy deposit in chromospheric layers (or probably higher) by non-thermal electrons or by thermal conduction. When the flare energy is transported toward lower layers, it can produce a local overpressure that drives both upward and downward mass motions. These motions can be detected through Doppler shift measurements in chromospheric and coronal lines. \cite{anto82,anto85}, \cite{anto83}, \cite{canf87}, \cite{zarr88}, \cite{wuls94}, \cite{ding96}, and \cite{dosc05} obtained blueshifts of 200--400~km~s$^{-1}$ in the Ca XIX line using the Bent and Bragg Crystal Spectrometer (BCS) onboard the Solar Maximum Mission (SMM; \citealt{acto80}) data and Yohkoh/BCS \citep{culh91} data, respectively. Similar measurements using data from the Coronal Diagnostic Spectrometer (CDS; \citealt{harr95})  onboard the Solar and Heliospheric Observatory (SOHO) revealed upflow velocities of 60--300~km~s$^{-1}$ in the Fe XIX line \citep{teri03,teri06,bros04,delz06,mill06a,mill06b}. On the other hand, in chromospheric and transition region lines, redshifts of about tens of km~s$^{-1}$ were observed, implying downward motions \citep{wuls94,czay99,teri03,teri06,bros03,kami05,delz06}. Recently, \cite{mill09} and \cite{chen10} measured Doppler velocities in multi-lines at flare regions to study the evaporation process using Hinode/EIS \citep{culh07} data . They detected both upflows and downflows in different emission lines.

Chromospheric evaporation can be classified into two types, the explosive evaporation and the gentle one \citep{fish85a,fish85b,fish85c,mill06a,mill06b,bros09}. When the lines formed in the upper chromosphere and transition region are red shifted and the hotter lines are blue shifted, the case is considered as an explosive evaporation. When all the lines appear to be blue shifted, it is considered as a gentle one. In the hydrodynamic simulations of a flaring atmosphere subject to thick-target electron beam heating, \cite{fish85a} found that an energy flux of about 10$^{10}$~ergs~cm$^{-2}$~s$^{-1}$ can be served as an effective threshold between the gentle and explosive evaporation. Note that in observations, explosive evaporation may appear in both large flares and microflares \citep{mill06a,bros09,vero10,bros10,chen10}.

Although many observational results support the evaporation model, there are still some inconsistencies between the observations and models. Some hot lines, such as Ca XIX, Fe XIX, and Fe XXIII, show a dominant stationary component with, however, a relatively weak blue-shifted component indicative of hundreds of km~s$^{-1}$ upflows \citep{anto82,ding96,mill06a,mill09}; while the flare dynamic  models predict that these lines should be mostly blue shifted. In the case of explosive evaporation, there is a conversion from redshifts to blueshifts. The models \citep{fish85c} predict that the downflows occur only at temperatures of $\le$ 1~MK, which was supported by many obsevations \citep{kami05,mill06a,delz06}. However, downflows were also detected at much higher temperatures, e.g., 2.0~MK \citep{mill08}. These pose challenges to the theoretical models.

With the high spatial and spectral resolution data of Hinode/EIS, we investigate in detail the process of chromospheric evaporation in a flaring region using multi-lines with different temperatures. Our work focuses on the two problems as shown above. Besides, we also find that there appear opposite line shifts in regions of different magnetic polarities. In the following, we describe the observations and data reduction in \S\ref{obs}. The results are shown in \S\ref{result}. We make a discussion in \S\ref{discussion}.

\section{Observations and data reduction}
\label{obs}

\begin{figure*}
\begin{center}
\includegraphics[height=7cm]{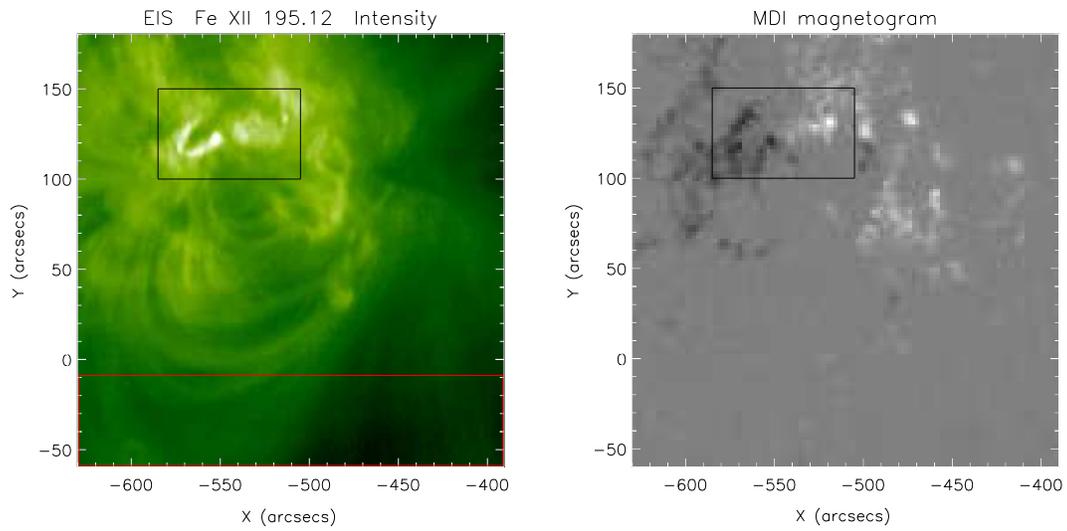}
\caption{Image of the Fe XII 195.12~\AA~intensity and MDI magnetogram for NOAA AR 10938. The 195.12~\AA~image is reconstructed from the second EIS scanning for the AR between 02:20:30 and 02:46:49~UT on 2007 January 16, which covers the impulsive phase of the flare. The black box refers to the flaring region that is studied in detail. The red box represents the quiet region that is used to establish reference wavelengths.}
\label{EISandMDI}
\end{center}
\end{figure*}

\begin{figure*}
\begin{center}
\includegraphics[height=15cm]{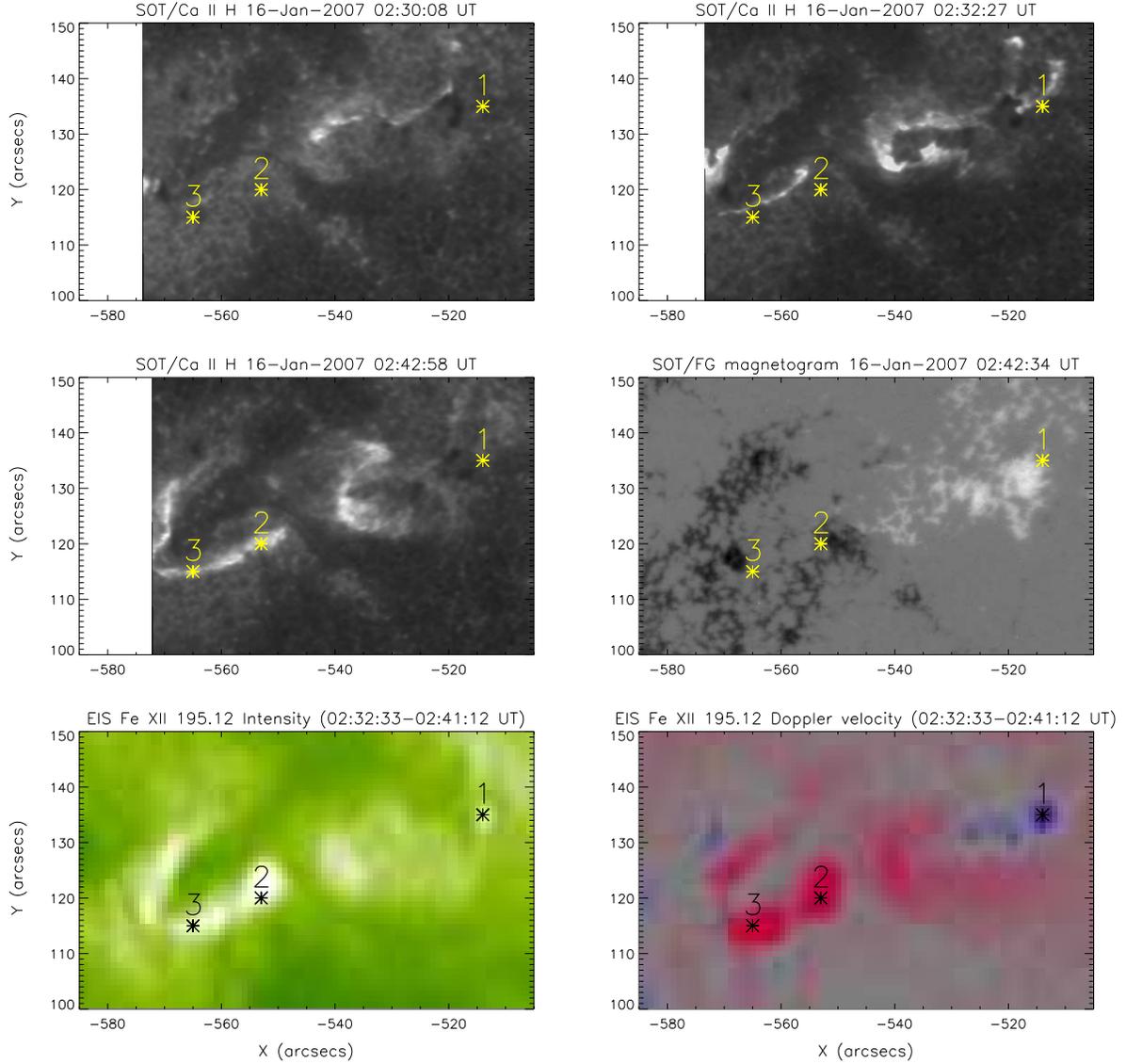}
\caption{Top and middle panels: SOT Ca II H images and SOT/FG longitudinal magnetogram. Bottom panels: EIS Fe XII 195.12~\AA~intensity and velocity maps. EIS scanned this region at the impulsive phase of the flare. The three points marked with 1, 2, and 3 are selected to derive the Doppler velocities in detail. The FOV of all the images here is the same as the black box shown in Figure~\ref{EISandMDI}.}
\label{box}
\end{center}
\end{figure*}

\begin{figure*}
\begin{center}
\includegraphics[height=13cm]{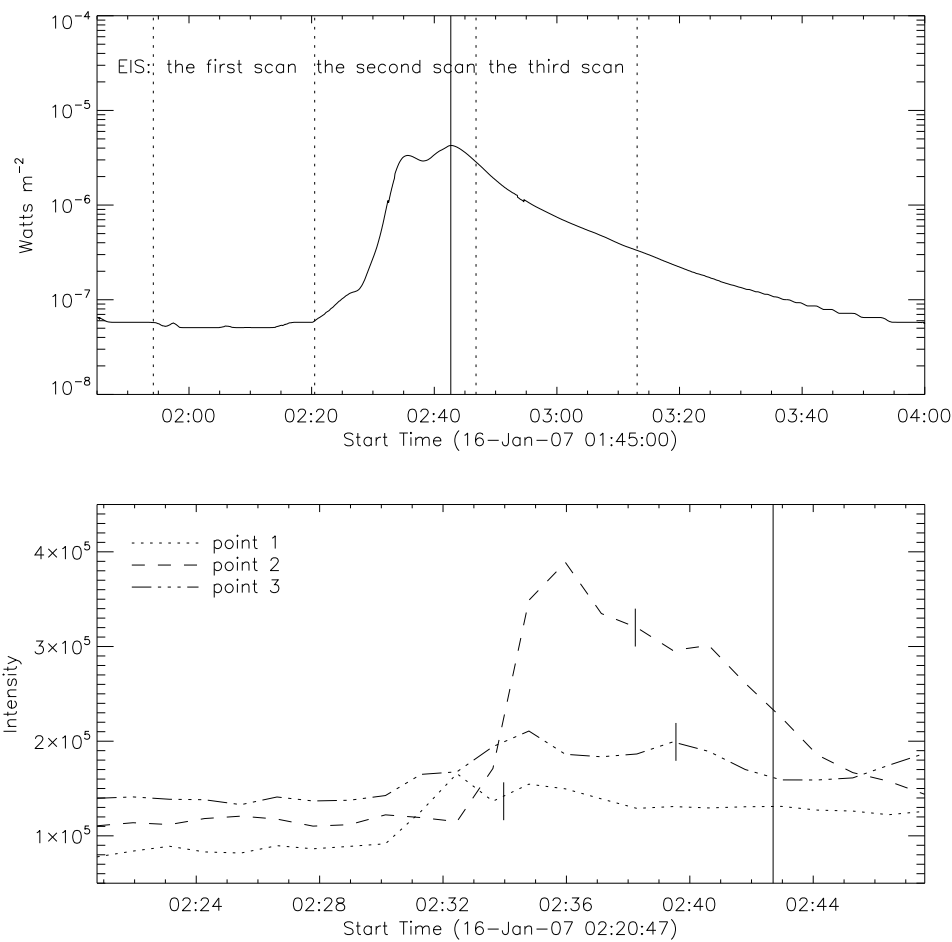}
\caption{Top panel: GOES 1--8 \AA~soft X-ray light curve of the flare. The vertical solid line indicates the flux peak, and the vertical dashed lines show the time ranges of the three EIS scans, which correspond to the preflare, impulsive, and post-impulsive phases of the flare, respectively. Bottom panel: SOT Ca II H intensity evolution at the three points in the impulsive phase. The long vertical line indicates the soft X-ray peak time at 02:42:42~UT, and the three short vertical lines represent the times at 02:33:58, 02:38:14, and 02:39:33~UT when EIS scanned over the three points, respectively.}
\label{flux_curve}
\end{center}
\end{figure*}

The observations presented here are for a GOES C4.2 class flare that started at 02:22 UT on 2007 January 16. The flare is located in the core of NOAA AR 10938. Figure~\ref{EISandMDI} shows the Hinode/EIS Fe XII 195~\AA~intensity map and the magnetogram measured by the Michelson Doppler Imager (MDI; \citealt{sche95}) onboard SOHO for the active region. This region presents a bipolar magnetic structure, which is favorable for production of flares. The 1$\arcsec$ slit of EIS was used to raster over the AR of 240$\arcsec \times$240$\arcsec$ with an exposure time of 5 s requiring a total duration of about 26 min. EIS scanned this region 3 times from 01:54:11 UT, corresponding to the preflare, impulsive, and post-impulsive phases of the flare. Our main interest focuses on  the second one between 02:20:30 and 02:46:49~UT that covers the impulsive phase. The area with a FOV of 80$\arcsec \times$50$\arcsec$, marked by the black box in Figure~\ref{EISandMDI}, contains the flare and is studied in detail here. Figure~\ref{box} shows the Ca II H images observed by the Solar Optical Telescope (SOT; \citealt{tsun08}) onboard Hinode, the SOT filter (FG) magnetogram, the Fe XII 195~\AA~intensity map, and the Doppler velocity of the Fe XII line for this region. The flare has two ribbons as seen from the Ca II H images. The right ribbon started to brighten at $\sim$02:30~UT; the left one appeared at $\sim$02:32~UT and disappeared later than the right one. In particular, we select three points showing significant upflows or downflows in most spectral lines for study. From the Ca II H images and the magnetogram, these points are located at the flare ribbons and close to strong magnetic patches. Point 1 lies in the positive polarity region and at the western part of the right ribbon, while points 2 and 3 lie in the negative polarity region and at the left ribbon. The time for this scan (the second one) is just before the GOES soft X-ray flux reached its maximum at 02:42~UT. Figure~\ref{flux_curve} shows the GOES 1--8 \AA~light curve of the flare and the SOT Ca II H intensity evolution at the three points in the impulsive phase. We also mark the EIS scanning time ranges and the time when EIS scanned over the three selected points. Note that the GOES light curve shows two peaks, which may correspond to brightenings of the two flare ribbons.


The EIS spectrum at each location in the raster contains 17 spectral lines. We select 11 lines among them spanning a temperature range of 0.05--16~MK. Details of the lines are shown in Table~\ref{line_data}. The majority of these lines are well resolved with no blends or only trivial blends that can be safely ignored in the active region \citep{youn07b}. We find that most of the observed line profiles have symmetric Gaussian shapes that can be fitted using a single Gaussian function. However, in some areas, the line profiles are asymmetric and can be well fitted by double Gaussian components.


We reduce the data using the standard EIS software data reduction package. This includes the correction of detector bias and dark current, as well as hot pixels and cosmic ray hits, resulting in absolute intensities in ergs~cm$^{-2}$~s$^{-1}$~sr$^{-1}$~\AA$^{-1}$. We also make a correction for a slight tilt of the slit on the CCDs. An additional effect that is corrected for is a variation of line positions over the Hinode orbit due to temperature variations in the spectrometer. Such an orbital variation is obtained by averaging the centroid positions over a length of slit for which the underlying solar region is mostly a quiet region. We choose the bottom 50 rows in the EIS Fe XII raster (see the red box in Figure~\ref{EISandMDI}) as a quiet region. The result is then subtracted from the line center positions measured in all wavelength windows.

EIS does not have an absolute wavelength calibration. We adopt the observed line centers averaged over the quiet region (the red box in Figure~\ref{EISandMDI}) as the rest wavelengths. Because of the EIS effective area that only peaks at about 195 and 275 \AA\ in the short wavelength (SW) and long wavelength (LW) bands, respectively, and the weak emission in the quiet region, we cannot obtain the accurate line centers for some lines, especially the high temperature lines. Therefore, we use the method by \cite{brow07} and the CHIANTI package \citep{dere97, dere09} to determine the reference wavelengths for those lines. To check the reliability of the methods, we use four strong lines, e.g., Fe XII, Fe XIII, Fe XIV, and Fe XV, for test. We find that the wavelengths of the line centers determined using the above methods are nearly the same with a deviation being within $\pm$0.003 \AA, which induces a velocity uncertainty of no larger than 5 km~s$^{-1}$.


We co-align the SOT/FG and EIS images by adopting the method described by \cite{guoy09}. Taking into account the instrumental offset between the images taken in the two EIS CCDs, we also shift the LW images by 2$\arcsec$ in the solar X direction and 17$\arcsec$ in the solar Y direction \citep{youn07a}.

\section{Results}
\label{result}

\subsection{Point 1: Upflows-dominated in the positive polarity region}
\label{upflow}

\begin{figure*}
\begin{center}
\includegraphics[height=7.5cm]{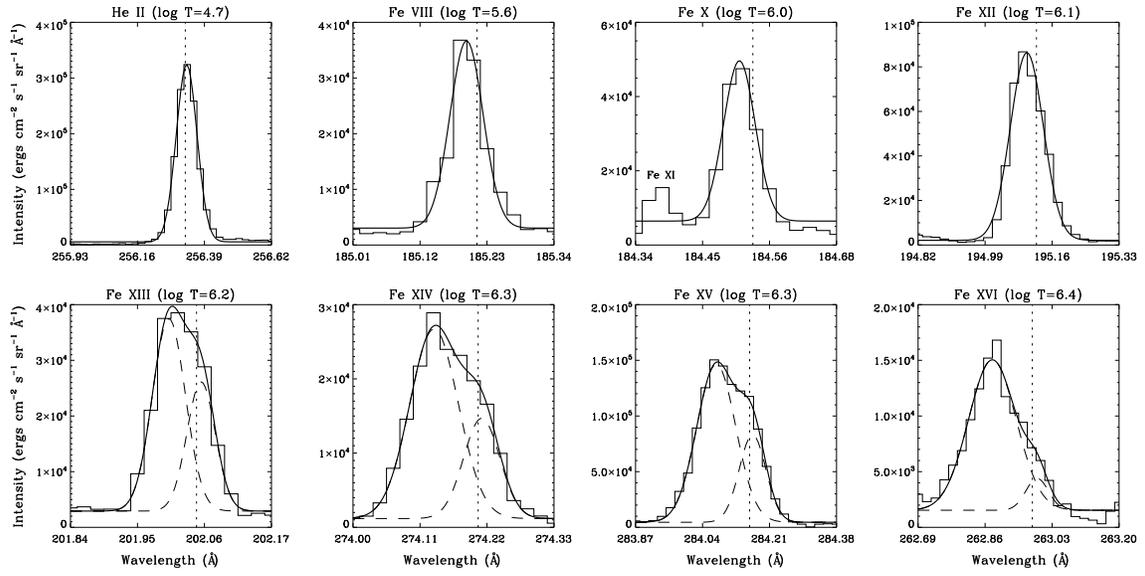}
\caption{Line profiles and fitting results for the 8 emission lines in the impulsive phase at point 1. The histograms are observed profiles and the solid lines are fitting results. The vertical dashed lines represent the reference wavelengths. The lines shown in the bottom row are well fitted by two Gaussian components that are plotted with dashed curves.}
\label{profilep1}
\end{center}
\end{figure*}

\begin{figure*}
\begin{center}
\includegraphics[height=11cm]{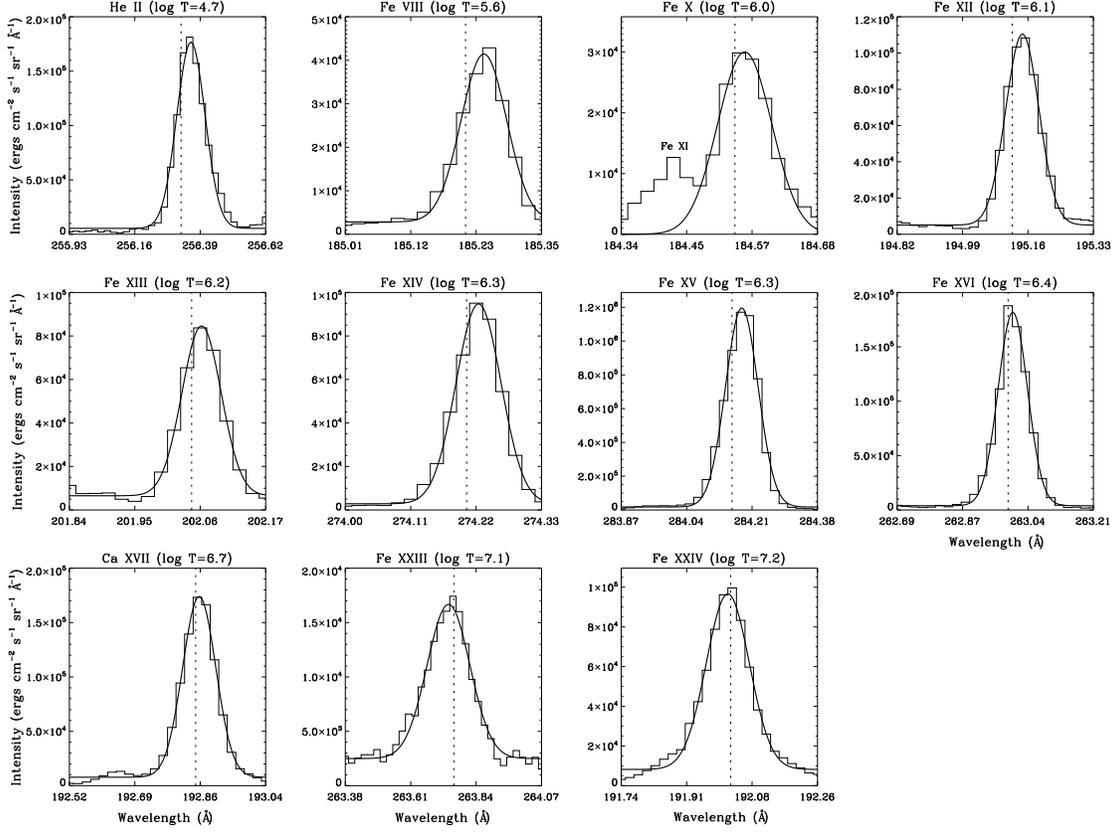}
\caption{Line profiles and fitting results for the 11 emission lines in the impulsive phase at point 2. The histograms are observed profiles and the solid lines are fitting results. The vertical dashed lines represent the reference wavelengths. All the line profiles are fitted with a single Gaussian function.}
\label{profilep2}
\end{center}
\end{figure*}

\begin{figure*}
\begin{center}
\includegraphics[height=7.5cm]{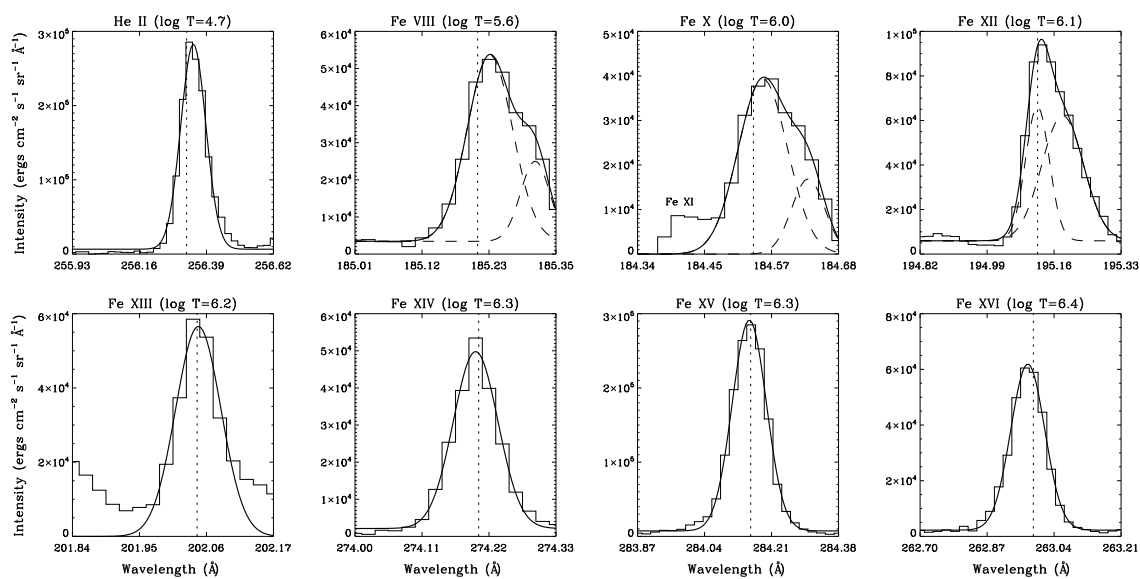}
\caption{Line profiles and fitting results for the 8 emission lines in the impulsive phase at point 3. The histograms are observed profiles and the solid lines are fitting results. The vertical dashed lines represent the reference wavelengths. Three lines (Fe VIII, Fe X, and Fe XII) are fitted by two Gaussian components that are plotted with dashed curves.}
\label{profilep3}
\end{center}
\end{figure*}

EIS scanned point 1 at 02:33:58 UT in the impulsive phase of the flare (see Figure~\ref{flux_curve}). This point shows upflows in all emission lines except for the He II line. Because of line blending we do not use the Ca XVII and Fe XXIV lines. We also ignore the Fe XXIII line since it is very weak in this region. We plot the line profiles and their fitting curves in Figure~\ref{profilep1}. Note that in the wavelength window of the Fe X 184.54 line, there exists the Fe XI 184.41 line; however, the latter is quite distinguishable from the former and therefore does not affect the fitting result (see also Figures~\ref{profilep2} and~\ref{profilep3}). Some of the line profiles are fitted by double Gaussian components. We also calculate the intensity ratio of the blue component to the stationary one. The ratio and the Doppler velocity for the blue component are listed in Table~\ref{velp1}. From the results we can see that most of the lines show obvious blueshifts. The upflow velocity increases with temperature from several tens of km~s$^{-1}$ to the highest one of 116 km~s$^{-1}$; the intensity ratio of the two components, derived from the double Gaussian fitting, has also an increasing tendency. For the Fe XIII line, the intensity of the blue component is just 1.72 times that of the stationary component. However, the ratio increases to 9.95 for the Fe XVI line. This indicates that the blue components are dominant over the stationary components for most of the lines, especially the lines with higher formation temperatures.


We also measure the average Doppler velocity over the 9 EIS pixels around point 1 in the preflare and post-impulsive phases and plot the value against the temperature in Figure~\ref{velp1p2}. The same is done for point 2. Note that the velocities corresponding to the blue components from the double Gaussian fitting are marked with the plus symbols. We find that most of the lines show significant blueshifts or blue-shifted components in the impulsive phase; while the shifts are trivial in the preflare and post-impulsive phases.


\subsection{Points 2 and 3: Downflows-dominated in the negative polarity region}
\label{downflow}

In contrast to the case of point 1, point 2 shows obvious downflows. EIS scanned point 2 at 02:38:14~UT, slightly before the GOES soft X-ray peak time at 02:42~UT. We fit all the line profiles at point 2 using a single Gaussian function (shown in Figure~\ref{profilep2}). Different from point 1, point 2 shows strong emission in high temperature lines (e.g., the unblended Fe XXIII line). \cite{youn07b} reported that the Ca XVII line completely dominates the other lines in large flares. Therefore, we can ignore the blendings of the Ca XVII line with the O V line and the Fe XI line. The blending of the Fe XXIV line with the Fe XI line can also be ignored for the same reason. The average Doppler velocity over the 9 pixels around point 2 is also plotted in Figure~\ref{velp1p2}. We find that most of the lines are red shifted, including the Fe XVI line (2.5~MK) and even the Ca XVII line (5.0~MK); the velocity decreases with temperature in the impulsive phase. Only the Fe XXIII and Fe XXIV lines show blueshifts with speeds of tens of km~s$^{-1}$. This means that the temperature division between upflows and downflows is about 5~MK. Note that in previous studies, the highest temperature division is 2~MK as reported by \cite{mill08}. Our result poses a challenge to the flare dynamic models.

For point 3, scanned by EIS at 02:39:33~UT, we also detect obvious downflows in most of the lines. It shows a similar behavior to point 2 through the entire flare process. One of the differences between them is the temperature division between upflows and downflows in the impulsive phase of the flare. It is about 2~MK at point 3. Another difference is that not all the line profiles at point 3 can be fitted with a single Gaussian function. Some line profiles are asymmetric that can be well fitted with two components (see Figure~\ref{profilep3}), especially for the low temperature lines. We notice that line blending can cause asymmetric line profiles, such as the possible blending of the Fe XII 195.12 and Fe XII 195.18 lines. However, the Fe X line is not blended with others. The blending in the Fe VIII line can be ignored at the flare ribbons \citep{youn07b}. Therefore, the apparent asymmetries, as well as the large widths, of the Fe X and Fe VIII lines imply that there exist multi-component downflows in the line formation layers \citep{dere84,ying09}.


\section{Discussions and conclusions}
\label{discussion}

We have presented the line profiles and Doppler velocities at flare ribbons of different magnetic polarities to study the chromospheric evaporation process. The key findings of our study are as follows: (1) The flare ribbons with different magnetic polarities can show different patterns of upflows or downflows in the impulsive phase of the flare; (2) the line profiles at one point of positive magnetic polarity are mostly blue shifted, with the blue-shifted component dominating over the stationary one for especially hotter lines; (3) downflows are detected at the points of negative magnetic polarity in lines of relatively high temperatures (up to 2.5--5.0~MK); (4) there exist multi-velocity components in some lines, either cooler or hotter, depending on the points in the flaring regions.

The upflows at point 1 in the positive magnetic polarity region are basically consistent with the scenario of a gentle evaporation process. The most significant finding here is the dominance of blue-shifted components in the line profiles, which is well consistent with the prediction of theoretical dynamic models. In previous observations, high temperature lines (e.g., Fe XIX, Ca XIX, and Fe XXIV) were found to be dominated by a stationary component \citep{anto82,ding96,mill06a,mill09}. It is difficult to explain such observational results in terms of the basic dynamic models. \cite{dosc05} argued that the stationary component is from the top of the flare loop where the evaporated mass is accumulated, or that the mass is moving perpendicular to the line of sight. They also suggested a possibility that the instrumentation is not sensitive enough to detect the earliest blueshifted emission; by the time the emission level has risen sufficiently, the flare loops have already been filled. \cite{fale09} took into account the geometrical dependence of the line-of-sight velocities of the plasma motions along the loops inclined toward the solar surface as well as a distribution of the flare sites over the solar disc. They concluded that the stationary component can be observed for all flares during their early phases of evolution; on the other side, the blue-shifted component may be undetectable even for plasma moving along the flaring loop with a very high velocity. In spite of these explanations, searching for a dominant blue-shifted component in high temperature lines during flares is still an interesting task to reduce the discrepancy between observations and models. In this work, although we cannot accurately determine the Doppler velocities of the Fe XXIII (12.5~MK) and Fe XXIV (15~MK) lines at point 1 because of the EIS sensitivity limitation and line blending, we confirm that some line profiles are dominated by blue-shifted components. Our result shows further that the intensity ratio of the blue-shifted component to the stationary one increases with temperature.

\begin{figure*}
\begin{center}
\includegraphics[height=7.5cm]{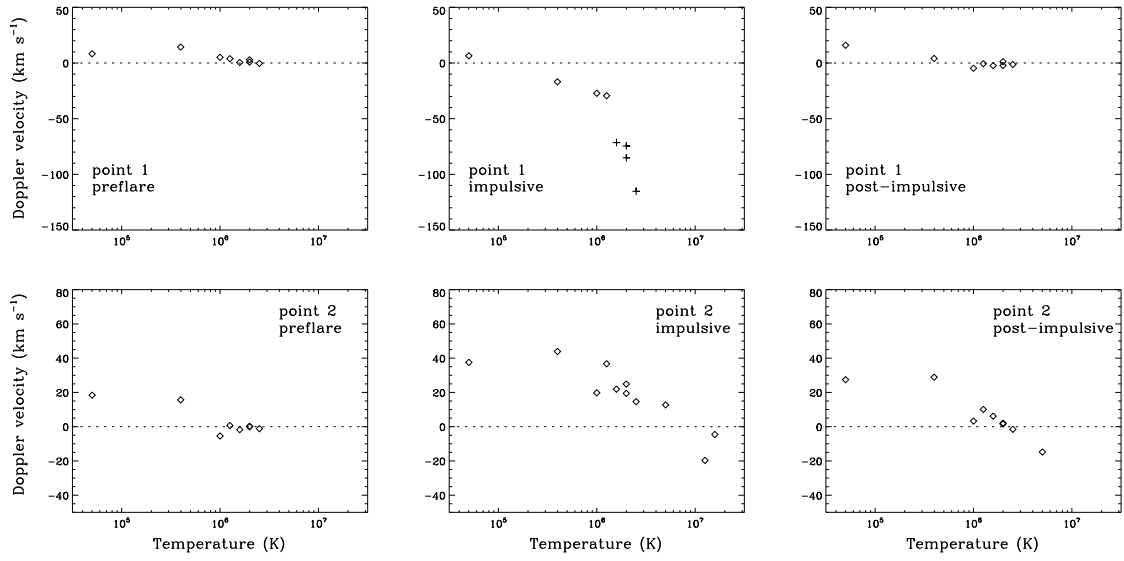}
\caption{Doppler velocities as a function of temperature in the three phases of the flare at points 1 and 2. Positive values represent redshifts and negative ones represent blueshifts. The diamonds refer to the velocities that are obtained from single Gaussian fitting, while the plus sign shows the velocities of the blue-shifted component from double Gaussian fitting.}
\label{velp1p2}
\end{center}
\end{figure*}

The counterpart of upflows is downflows in the flaring loop subject to a rapid energy deposition and momentum balance. Downflows were mostly detected in chromospheric lines before the SOHO era (e.g., \citealt{ichi84,ding95}). This is known as the phenomenon of chromospheric condensation \citep{fish89}. With space instruments, downflows are also seen in some coronal lines. Most previous observations showed that the downflows occur in temperatures of $\le$~1~MK \citep{kami05,mill06a,delz06}. Quite recently, \cite{mill09} detected downflows at a temperature of 1.5~MK. Moreover, \cite{mill08} reported a redshift of 14 km~s$^{-1}$~in the Fe XV line (2~MK) at flare footpoints. The fact that the downflows exist in such a high temperature provides a constraint to the dynamic model. Our measurements presented here show that redshifts can appear in lines of even higher temperatures (Fe XVI, Ca XVII) at the points in the negative magnetic polarity region. Most of the line profiles at point 2 show redshifts while only very few are blue shifted. This case can be regarded as an explosive evaporation \citep{mill06a,bros09}. If we take the division temperature between upflows and downflows as the site of energy deposition by the electron beam, appearance of the unusually high division temperature at point 2 implies a rather high energy deposition site. A possible reason is that the flare loop has been filled with enough mass before the impulsive phase. A high coronal density can prevent the electron beam from penetrating deeper. Here we do not detect blueshifts at temperatures higher than 15 MK due to the limitation of the EIS dynamic range, but we find that the Fe XXIII and Fe XXIV line intensities at this point are still strong and their profiles are well Gaussian-shaped in this event. This seems to support the hypothesis of a high coronal density. In addition, \cite{bros03} detected redshifted emission that persisted at least 20 min after the cessation of blueshifts and suggested that flare plasma, heated and accelerated upward during the impulsive phase, subsequently cooled and fell back down in what may be thought of as ``warm rain''. For the present event, scanning the active region by EIS lasted about 26 min; therefore, we do not have an enough high temporal resolution here. However, from Figure~\ref{velp1p2}, we can find that the downflows still exist in the post-impulsive phase at point 2, contrary to the blueshifts at point 1, which are diminished to nearly zero in the post-impulsive phase. From the SOT Ca II H movie, we find that the flare ribbons are not brightened simultaneously. Therefore, we cannot exclude the warm rain as a possible cause of the downflows.

The three points discussed here are located in different magnetic polarity regions. They show different patterns of mass flows. Point 1 lies in the positive polarity region while points 2 and 3 in the negative polarity region. We check further the magnetic connectivities between the positive and negative magnetic polarities. The available data cannot provide a definite conclusion regarding whether point 1 is magnetically connected to point 2 or 3. However, even if they belong to different magnetic loops, the different evaporation patterns shown at them suggest that in one flaring region, the heating mechanisms and atmospheric conditions may vary from point to point.

\acknowledgments
The authors would like to thank the referee for valuable comments on the paper and Feng Chen, Xin Cheng, and Zongjun Ning for helpful discussions. This work was supported by NSFC under grants 10828306 and 10933003 and by NKBRSF under grant 2011CB811402. Hinode is a Japanese mission developed and launched by ISAS/JAXA, collaborating with NAOJ as a domestic partner, and NASA (USA) and STFC (UK) as international partners. Scientific operation of the Hinode mission is conducted by the Hinode science team organized at ISAS/JAXA. Support for the post-launch operation is provided by JAXA and NAOJ (Japan), STFC (U.K.), NASA, ESA, and NSC (Norway).

\bibliographystyle{apj}

\begin{table}
\begin{center}
\small
\caption{\textsc{\small{Emission lines used in this study}} \small{\citep{delz10}}}
\label{line_data}
\begin{tabular}{lcc}
\tableline
\tableline
\multicolumn{1}{l}{Ion}	&$\lambda$(\AA) &$T_{max}$~(MK) \\
\tableline
\ion{He}{2}	    &256.32	&0.05   \\
\ion{Fe}{8}	    &185.21	&0.4    \\
\ion{Fe}{10}	&184.54	&1.0	\\
\ion{Fe}{12}	&195.12	&1.25	\\
\ion{Fe}{13}	&202.04 &1.5	\\
\ion{Fe}{14}	&274.20	&2.0	\\
\ion{Fe}{15}	&284.16	&2.0	\\
\ion{Fe}{16}	&262.98	&2.5	\\
\ion{Ca}{17}	&192.82	&5.0	\\
\ion{Fe}{23}	&263.76	&12.5	\\
\ion{Fe}{24}	&192.03	&16.0	\\
\tableline
\normalsize
\end{tabular}
\end{center}
\end{table}

\begin{table}
\begin{center}
\small
\caption{\textsc{\small{Velocity of the blue-shifted component and its intensity ratio to the stationary component at point 1}}}
\label{velp1}
\begin{tabular}{lcc}
\tableline
\tableline
\multicolumn{1}{l}{Ion}	&$V_d$(km~s$^{-1}$) &ratio \\
\tableline
\ion{He}{2}	    &+5	    &   \\
\ion{Fe}{8}	    &-28	&   \\
\ion{Fe}{10}	&-36	&	\\
\ion{Fe}{12}	&-37	&	\\
\ion{Fe}{13}	&-68    &1.72	\\
\ion{Fe}{14}	&-80	&2.53	\\
\ion{Fe}{15}	&-89	&2.64	\\
\ion{Fe}{16}	&-116	&9.95	\\
\tableline
\normalsize
\end{tabular}
\end{center}
\end{table}

\end{document}